\documentclass[12pt]{iopart}

\usepackage{epsfig}
\usepackage{graphicx}
\usepackage{bm}
\usepackage{iopams}
\usepackage{lscape}
\usepackage[usenames,dvipsnames]{color}

\newcommand{\lfs}{f_{028}}
\newcommand{\eis}{e_{IS}}
\newcommand{\cna}{\{1,4,2\}}

\newcommand{\ave}[1]{\langle {#1} \rangle_0}

\newcommand{\CC}{\mathcal{C}}
\newcommand{\tobs}{t_{\rm obs}}

\begin{document}

\title{Structure of inactive states of a binary Lennard-Jones mixture}

\author{Daniele Coslovich$^1$ and Robert L. Jack$^2$}
\address{$^1$ Laboratoire Charles Coulomb, UMR 5221 CNRS-Universit\'e de Montpellier, Montpellier, France}
\address{$^2$Department of Physics, University of Bath, Bath BA2 7AY, United Kingdom}
\ead{daniele.coslovich@umontpellier.fr}

\begin{abstract}

We study the structure of inactive states in a prototypical model glass, the Kob-Andersen binary Lennard-Jones mixture. These inactive states are obtained by transition path sampling and are at dynamical phase coexistence with an active equilibrium state. Configurations in the inactive states are kinetically stable and are located in deeper basins of the energy landscape than their active counterparts. 
By analyzing trajectory-to-trajectory fluctuations within the inactive state, we assess correlations between kinetic stability, energy and other structural properties.
We show that measures of local order associated to stable local packings and bond-orientational order are weakly correlated with energy and kinetic stability.
We discuss what kinds of structural measurement might capture the relevant dynamical features of the inactive state.

\end{abstract}
\submitto{JSTAT}

\maketitle

\section{Introduction}

How should we characterize the amorphous structure of glassy materials?  This is a much debated question, which has given rise to a variety of theoretical and experimental proposals~\cite{Tarjus2005,Lubchenko2007,Chandler2010,Kurchan2011}. One idea, which appears in many different theories, is that the large relaxation times that are found near to glass transitions are associated with some universal form of co-operative behavior characterized by a growing length scale~\cite{Garrahan2002b,Bouchaud2004,Tarjus2005,Kurchan2011}.  Alternatively, one might seek a local explanation for glassy behavior: one can attribute a large time scale to a population of particles that find themselves in stable local environments, requiring a large activation energy in order to move~\cite{Frank-PRS1952,Ma_2015}.
These two explanations are different, but they are not necessarily contradictory: it may well be that stable local environments predominate in glassy materials, and that the length scales for co-operative motion also grow.  
Moreover, complex arrangements of stable local motifs can give rise to so-called medium range order~\cite{Tanaka_2012,Ma_2015}, which extends over length scales larger than the typical nearest neighbor distance.

Evidence that both local and cooperative mechanisms can coexist in glassy systems comes from several recent simulation studies.
Some of them have focused on simple mixtures of Lennard-Jones particles, which accumulate a significant amount of local order upon cooling~\cite{Coslovich-JCP2007,Coslovich-PRE2011,Malins_Eggers_Royall_Williams_Tanaka_2013,Malins_Eggers_Tanaka_Royall_2014}.
The local structure of these systems is characterized by stable motifs, called locally favored structures (LFS), whose abundance increases markedly as temperature is lowered.
Although the specific symmetry of such local structures is system-dependent, the mechanism seems to be rather general and carries over to metallic glasses~\cite{Cheng_Ma_Sheng_2009,Hirata_Guan_Fujita_Hirotsu_Inoue_Yavari_Sakurai_Chen_2011,Ding_Cheng_Sheng_Ma_2012,Soklaski_Nussinov_Markow_Kelton_Yang_2013}.
Other studies have focused on more complex forms of ``amorphous order'', such as point-to-set correlations~\cite{Biroli_Bouchaud_Cavagna_Grigera_Verrocchio_2008,Berthier-static-PRE2012,Hocky-PRL2012,Fullerton2014,BerthierJack2015}, which may be expected to govern the co-operative relaxation mechanisms at low temperature.
In the end, the question of the most appropriate description of the amorphous structure depends strongly on the extent to which different measurements can be used to predict the behavior of glassy systems~\cite{Berthier-PRE2007,Jack2014}. Of course, the answer to this question also depends on the specific system being studied~\cite{Hocky2014}.

A strong indication of universal phenomenology in glassy materials is the existence of phase transitions, which are associated with diverging length scales and characteristic order parameter fluctuations~\cite{Chandler2010,Franz-EPJE2011,Bouchaud2004}.  Such transitions may occur at so-called ideal glass transitions, or they may occur in response to external perturbations such as random pinning~\cite{Cammarota-PNAS2012}, dynamical biasing~\cite{Hedges2009}, or in systems of coupled replicas~\cite{Franz1997,BerthierJack2015}.  Here we consider a well-studied binary mixture of Lennard-Jones particles, originally proposed by Kob and Andersen (KA)~\cite{kamix-pre95-1,kamix-pre95-2}.  We focus on systems that are biased dynamically so that their particles move less than is typical at equilibrium~\cite{Hedges2009}.  The specific biasing procedure used is based on the mathematical theory of large deviations~\cite{Touchette2009}.  It leads to a dynamical phase transition at which structural relaxation of the system appears to stop completely, as the system enters an inactive state.  Such phase transitions were predicted on the basis of dynamical facilitation theory~\cite{Garrahan2007}: in fact they occur in a variety of glassy model systems~\cite{Hedges2009,Speck2012b,Jack2010,Turner2015}, consistent with predictions of universality.

As the dynamical bias is applied to the KA mixture, the system is maintained in contact with a heat bath at temperature $T$.  However, the structure of the system changes significantly at the phase transition, so the inactive state differs strongly from equilibrium states at that temperature~\cite{Jack2011,Speck2012b}.  The inactive states are also kinetically stable, in that they take an unusually long time to recover back to equilibrium when the dynamical bias is removed~\cite{Jack2011,Speck2012b}. 

Here, we analyze the local structure of the inactive states, to measure the extent to which stable local environments determine the properties of these stable glasses. We achieve this by analyzing sample-to-sample fluctuations among a set of stable glass configurations.  We measure correlations between stabilities, energies, locally favored structures, crystalline order and local composition. We find that configurations with the lowest energies tend to be the most stable, but neither their stability nor their low energy can be attributed to a single structural motif. Our conclusion is that the correlations between structure, energy and stability of amorphous states in this model cannot be explained on the basis of a single locally favored structure.

\section{Numerical Methods}
\label{sec:methods}

\subsection{KA mixture}

The KA mixture consists of particles of two species, $A$ and $B$, with Lennard-Jones interactions, as described in~\cite{kamix-pre95-1,kamix-pre95-2}.  Species $A$ is larger and $80\%$ of the particles are of this type.  The units of length and energy are set by by the interactions between $A$-particles, via parameters $\sigma=\sigma_{\rm AA}=1$ and $\epsilon=\epsilon_{\rm AA}=1$ respectively.  When considering large deviations of the dynamical activity, it is convenient to use overdamped (Brownian) dynamics, which we implement using the Monte Carlo (MC) method of~\cite{BerthierKob2007}.  Within this scheme, the (bare) diffusion constant of a single free particle is $D_0=a_0^2/(6\tau_{\rm MC})$ where $a_0=0.075\sigma$ is the maximal MC step in each Cartesian direction and $\tau_{\rm MC}$ is the time associated with a single Monte Carlo sweep (one attempted move per particle).  The natural physical time scale for this system is $\Delta t = \sigma^2 / D_0$, which is of the order of the Brownian time.  This leads to $\Delta t =6 \tau_{\rm MC}/a_0^2$, corresponding to approximately 1070 MC sweeps.  There are $N$ particles, and the position of particle $i$ at time $t$ is $\bm{r}_i(t)$.  The density is $\rho=1.2\sigma^{-3}$.

\subsection{Biased ensembles of trajectories}\label{sec:bias}

This article is concerned with the structure of inactive states, which are obtained by a dynamical biasing scheme, based on large deviation theory.  We follow the method
of~\cite{Hedges2009} (see also~\cite{Speck2012b,Fullerton2013}).  We consider trajectories of the system, which run from initial time $t=0$ to final time $t=t_{\rm obs}$. Large deviation theory is relevant in the limit $\tobs\to\infty$:  for numerical purposes, we take $\tobs \gg \tau_\alpha$ and use finite-size scaling methods to analyze the dependence of our results on $\tobs$~\cite{Hedges2009,Speck2012b}.  (Here $\tau_\alpha$ is the structural relaxation time.) The inactive states considered were obtained at temperature $T=0.6$, for which $\tau_\alpha \approx 12 \Delta t$.  

As in~\cite{Hedges2009}, we define the activity of a trajectory of the system as 
\begin{equation}
K = \Delta t \sum_{j=1}^{M} \sum_{i=1}^{N_A} | \hat{\bm{r}}_i(t_{j}) - \hat{\bm{r}}_i(t_{j-1}) |^2  ,
\end{equation}
where the times $(t_0,t_1,t_2,\dots,t_M)$ are equally spaced along the trajectory, as $t_j = j\Delta t$, with $M=\tobs / \Delta t$.  The sum over particles $i$ is restricted to particles of type $A$.  Also the position $ \hat{\bm{r}}_i(t) = \bm{r}_i(t) - \overline{\bm{r}}(t)$ is defined by subtracting the center of mass  $\overline{\bm{r}}(t)$ so that the activity $K$ does not couple to bulk translational motion, which can be significant in the relatively small systems considered here.

We denote averages in the equilibrium state of this system by $\langle \cdot \rangle_0$.  Then introduce a dynamical field $s$ which biases the system to low activity.  The effect of this field is similar to the effect of temperature changes in statistical mechanics: by analogy with the canonical ensemble, the average of some $O$ 
in the biased system (or ``$s$-ensemble'') is
\begin{equation}
\langle O \rangle_s =  \langle O {\rm e}^{-sK} \rangle_0 \frac{1}{Z(s,\tobs)},
\label{equ:s-ens}
\end{equation}
where $Z(s,\tobs) = \langle {\rm e}^{-sK}  \rangle_0$ is a dynamical partition function.  For the observable $O$, one might take $O=K$, or $O=E(t)$, the energy of the system at some specific time $t$ (with $0\leq t\leq\tobs$).  For $s>0$, one sees that (\ref{equ:s-ens}) assigns an increased statistical weight to trajectories with low activity $K$.  (In a similar way, the canonical ensemble of statistical mechanics assigns an increased weight to configurations with low energy.)  The result is that the $s$-ensemble biases trajectories to low activity, but without any direct bias on the structures that the system should adopt in order to achieve these inactive states.

Numerically, we use transition path sampling (TPS)~\cite{Bolhuis2002} to generate trajectories that are representative of the $s$-ensemble, so that averages of the form of (\ref{equ:s-ens}) correspond to averages over our sampled trajectories.  See~\cite{Fullerton2013} for an outline of the method used, which generates trajectories (sample paths) of the system according to a probability distribution
\begin{equation}
P[X|s] =  P[X|0] \cdot \frac{{\rm e}^{-s K[X]} }{ Z(s,\tobs) }.
\label{equ:PXs}
\end{equation}
Here $P[X|0]$ is the probability that trajectory $X$ occurs in the equilibrium state ($s=0$) and $P[X|s] $ is the probability of that trajectory in the $s$-ensemble.

In the following we consider data from $s$-ensembles sampled at $T=0.6$, with $N=256$ and $\tobs=250\Delta t$.  (There are $N_{\rm A}=205$ particles of type $A$ and $N_{\rm B}=51$ of type $B$.)  For these parameters, the activity $k(s) =  \langle K \rangle_s / (N_A\tobs)$ exhibits a crossover from active to inactive behavior at a field $s=s^* \approx 0.020$.  At $s=s^*$, the distribution of the activity $K$ is bimodal, with the two peaks corresponding to active (large-$K$) and inactive (small-$K$) states.  The inactive state that we consider is sampled at $s=s^*$, with the restriction to trajectories with $K/(N_A\tobs)< 0.039\sigma^2$.  This corresponds to sampling from the inactive phase, at the phase coexistence point.  As in~\cite{Hedges2009}, we use a criterion based on a common neighbor analysis to avoid crystallization of the system: details are given in Section~\ref{sec:structure} below.

We use two methods to improve the performance of the TPS method.  These are generalizations of standard methods for equilibrium sampling, but now applied to trajectories.  We use parallel tempering (see also~\cite{Speck2012b}) with 8 replicas, of which 6 use $s$-values very close to the estimated phase transition point $s^*$, while the remaining two use smaller $s$, where acceptance rates TPS moves are high.  In particular it is useful to have one replica with $s=0$, for which all TPS moves are accepted: this facilitates rapid exploration of trajectory space.  In some cases we also apply a bias potential $w(K)$ so that we sample from a distribution $P[X|s,w] \propto P[X|s] {\rm e}^{w(K)}$.   This approach facilitates sampling in cases where $P[X|s]$ contains two phases with a significant barrier between them (as happens at the first-order phase transition): in that case $w(K)$ can be chosen to enhance the likelihood of the system crossing the barrier between the phases.  Results for the relevant case $w=0$ can be readily obtained by standard histogram reweighting~\cite{frenkelsmit}.

In Section~\ref{sec:results}, we analyze various indicators of stability and local order of inactive states.  For local structural measurements, we consider averages over the full ensemble of trajectories generated by TPS.  However, when it is computationally expensive to measure the relevant quantities, we use a representative sample containing 38 trajectories.  Where we show scatter plots, the points shown should be interpreted as representative samples from $P[X|s]$.

\begin{figure}[t]
  \centering
\includegraphics[width=0.7\linewidth]{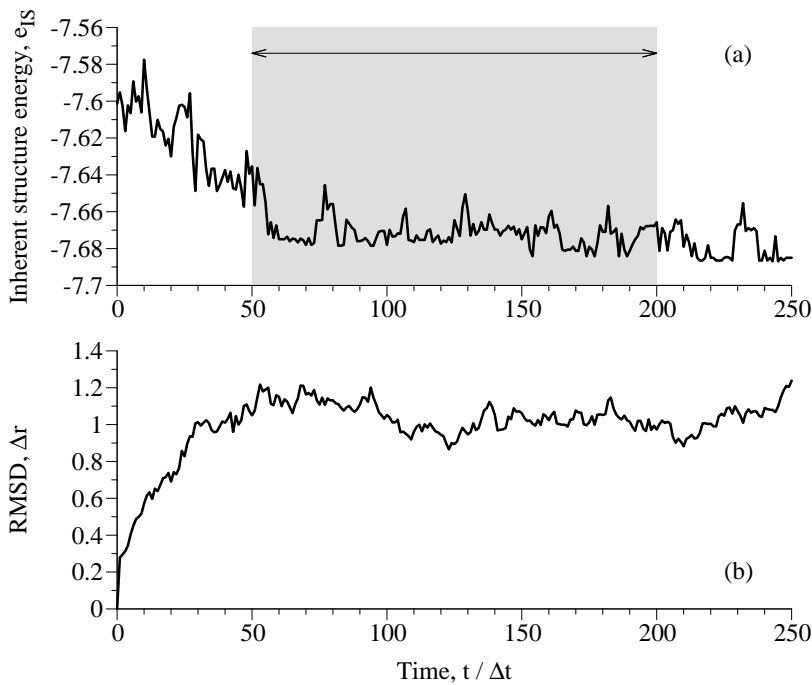}
\caption{\label{fig:window} Illustration of initial and final transient regimes for a representative trajectory with $K\approx7.77$.  In our analysis, we average the inherent structure energy, panel (a), over the central part of the trajectory (shaded area).  This avoids effects from transients near the initial and final times. Note that in this illustrative trajectory there is an initial transient but no final transient, but since the ensemble is time-reversal symmetric so the opposite situation is equally likely. Panel (b) shows the root mean square displacement $\Delta r = [(1/N)\sum_i |\bm{r}_i(t)-\bm{r}_i(0)|^2]^{1/2}$ over the same time period, showing that the particles move significantly only within the transient regime.}
\end{figure}

We note that the $s$-ensemble is not perfectly time-translation invariant~\cite{Garrahan2009,JackSollich2010}: there are transient regimes associated with the beginning and end of the trajectories.  However, time-translation invariance does hold in the ``bulk'' of the trajectories (that is, for times $t\gg\tau$ and $\tobs-t\gg\tau$, where $\tau$ is the time scale associated with decay of the transient).  For these reasons, when selecting representative inactive configurations from the $s$-ensemble, we take one configuration per trajectory, evaluated at time $t=\tobs/2$.  In some cases, we also take time-averages within the inactive state, to reduce the effect of intra-state fluctuations due to fast degrees of freedom.  In this case, the averages are taken over the time period $(\tobs/4) < t < (4\tobs/5)$.  Figure~\ref{fig:window} illustrates this procedure.

\subsection{Kinetic stability of the inactive state: definition of $t_{\rm melt}$} \label{sec:melt}

As in previous work~\cite{Jack2011,Speck2012b}, we measured the time it takes for inactive states to relax to equilibrium (or ``melt''), once the bias is removed.  We take the the central  configuration ($t=\tobs/2$) from each inactive trajectory and we run 100 independent dynamical trajectories from this configuration, at $T=0.6$.  For each initial configuration, we measure the average (time-dependent) energy $E(t)$: it fits well to an exponential function $E=E_{\rm eq} + (E_0 - E_{\rm eq}){\rm e}^{-t/t_{\rm melt}}$, from which we obtain a (configuration-dependent) ``melting time'' $t_{\rm melt}$.  (Here, $E_{\rm eq}$ is the equilibrium average energy at $T=0.6$, while $E_0$ and $t_{\rm melt}$ are fitting parameters.)

\subsection{Local order: definitions of structural measurements}\label{sec:structure}

We now describe our analyses of local structure.  To reveal the most important structural features of inactive states, we focus on inherent structures (IS), in which thermal distortions are removed.  For any configuration $\{\bm{r}_i(t)\}$, the inherent structure corresponds to the closest local minimum of the potential energy surface, which we determine using the  limited-memory Broyden-Fletcher-Goldfarb-Shanno (LBFGS) minimization algorithm~\cite{liu__1989}.  Because we truncate the Lennard-Jones interactions at $r_{c}=2.5\sigma_{\alpha\beta}$, a small fraction of energy minimizations does not converge to a strict local minimum, but to configurations that contain a spurious unstable mode.  These configurations are nonetheless retained in our inherent structure analysis since they are statistically indistinguishable from actual local minima.

\begin{figure}[t]
  \centering
  \begin{tabular}{llll}
  \includegraphics[width=0.22\linewidth]{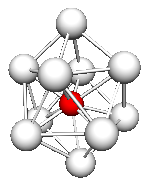} &
  \includegraphics[width=0.24\linewidth]{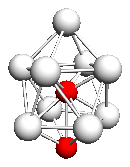} &
  \includegraphics[width=0.24\linewidth]{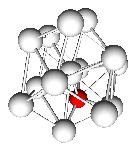} & 
  \includegraphics[width=0.24\linewidth]{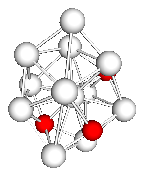} \\[-2.5em]
  {\footnotesize(a)} &  {\footnotesize(b)} &  {\footnotesize(c)} & {\footnotesize(d)} \\[1em]
  \end{tabular}
  \caption{\label{fig:snaps}Typical local structures associated to (0,2,8) Voronoi cells [panels (a) and (b)] and to (0,4,4,6) cells [panels (c) and (d)]. The former are identified as the LFS of the model, the latter are associated to fcc ordering. White and red spheres depict A and B particles, respectively. The majority of LFS have chemical coordination as in (a), while panel (b) displays an LFS with two B-particles along the symmetry axis. Note that (0,4,4,6) cells do not always always clear 6-fold symmetry, see panel (d). Bonds drawn according to a fixed cut-off distance, as explained in the main text.
  }
\end{figure}

Locally favored structures are natural candidates to explain the increased stability of inactive states sampled in the $s$-ensemble. LFS are identified
through a radical Voronoi tessellation of the inherent structures ~\cite{Gellatly_Finney_1982} using the Voro++ library~\cite{voro++}. At equilibrium, we focus on Voronoi cells centered on  $B$-particles, which possess 8 pentagonal faces and 2 quadrilateral faces. Such cells increase in number as temperature drops~\cite{Coslovich-JCP2007} and are identified as the preferred local structure of the KA mixture.  In the Voronoi nomenclature, these structures are referred to as $(0,2,8)$ cells~\footnote{The ``signature'' of a Voronoi cell $(n_3,n_4,n_5,\dots)$ where $n_k$ is the multiplicity of faces with $k$ vertices.}.  Note that, in contrast to other studies~\cite{Speck2012a,Malins_Eggers_Royall_Williams_Tanaka_2013,Malins_Eggers_Tanaka_Royall_2014}, we measure the overall fraction $f_{(0,2,8)}$ of LFS \textit{centers} instead of the concentration of particles being part of such LFS structures.  In addition, we monitor the fraction of $(0,4,4,6)$ cells centered around $A$-particles (see Figure~\ref{fig:snaps}c), which are associated with fcc and bcc crystal structures~\cite{tanemura_geometrical_1977}.  

To ensure that our results do not depend qualitatively on the technical details of the tessellation, we also performed a modified tessellation~\cite{DellaValle_Gazzillo_Frattini_Pastore_1994} to account for the different sizes of the particles in the KA mixture. In this case, segments connecting neighboring particles are bisected at a fraction $f_{\alpha\beta}=\sigma_{\alpha\alpha}/(\sigma_{\alpha\alpha}+\sigma_{\beta\beta})$ depending on the species of the particles' pair~\cite{DellaValle_Gazzillo_Frattini_Pastore_1994,Coslovich-JCP2007}. 
As is well known~\cite{Gellatly_Finney_1982}, there can be appreciable discrepancies between Voronoi cell statistics obtained using these two approaches. We found that, on average, signatures match about 80\% of the times and that $(0,2,8)$ cells are slightly less frequent when using the radical tessellation. These discrepancies do not affect, however, the main conclusions of our work, see Section~\ref{sec:results}.

\begin{figure}[t]
  \centering
\includegraphics[width=0.75\linewidth]{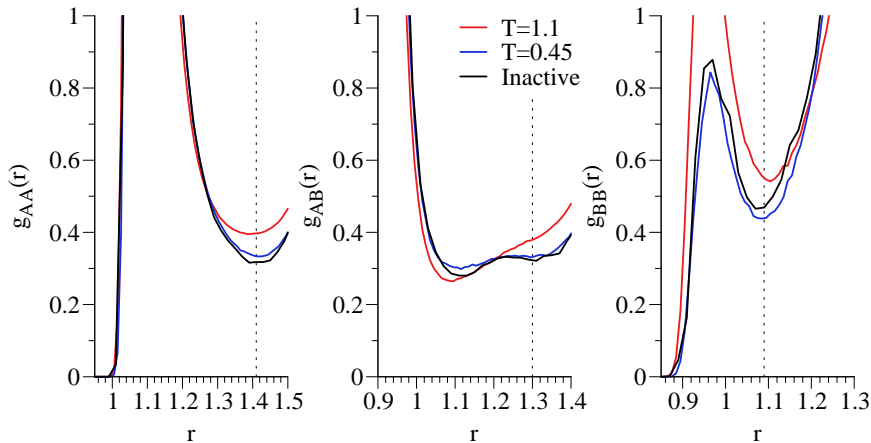}
\caption{\label{fig:gr} Partial radial distribution functions $g_{AA}(r)$, $g_{AB}(r)$, and $g_{BB}(r)$ for inherent structures at $T=1.1$, 0.45 and in the active state at $T=0.6, s=s^*$. The vertical dotted lines are located at $r=1.41$, 1.30, and 1.09, respectively.}
\end{figure}

Two additional structural metrics are employed to detect local packings with crystal-like order: bond-orientational order (BOO) parameters and common neighbor analysis (CNA). In both cases, our analysis relies on the notion of a particle's neighborhood, which we define as follows: two particles $i$ and $j$ are considered as neighbors, i.e. they form a bond, if their distance $r_{ij}$ is less than a threshold value $r_{\alpha\beta}^m$, which depends on the species $\alpha$ and $\beta$ of the particles. For inherent structures, the values of $r_{\alpha\beta}^m$ are 1.41, 1.30 and 1.09 for AA, AB and BB pairs, respectively. These values match the location of the first minimum in the relevant radial distribution function $g_{\alpha\beta}(r)$. We note that we choose $r_{AB}$  slightly larger than the location of the first minimum of $g_{AB}(r)$, to try to account for typical $A-B$ separations along the principal axis of the LFS, see Figure~\ref{fig:snaps}a.  Also note that the positions of these minima are rather insensitive to changes in temperature or activity, see Figure~\ref{fig:gr}.

To quantify the degree of local bond-orientational order in the system, we use a standard expansion of the bonds formed by neighboring particles in terms of spherical harmonics $Y_{lm}$~\cite{Steinhardt-PRL1981}.  For each particle $i$, the complex vector 
\begin{equation}
Q_{lm} = \sum_{j=1}^{N_b(i)} Y_{lm}(\hat{\bm{r}}_{ij})
\end{equation}
encodes information about the $l$-fold orientational symmetry of bonds formed by particle $i$ with its $N_b(i)$ neighbors. Here, $\hat{\bm{r}}_{ij}$ is the unit vector between particles $i$ and $j$.  From these quantities we then construct a set of rotationally-invariant bond-order parameters
\begin{equation}
  Q_l(i) = \sqrt{\frac{4\pi}{2l+1}\sum_{m=-l}^{l} |Q_{lm}(i)|^2}
\end{equation}
which are sensitive to different kind of symmetries of the local structure. Specifically, the $Q_6$ parameter is large for fcc and hcp structures and will be used in the following as a simple measure of local crystal-like order. \footnote{Note that although large $Q_6$ values can also correspond to strong local icosahedral order, our Voronoi tessellation shows that icosahedral structures are very scarce in the system. At low $T$, the fraction of $(0,0,12)$ Voronoi cells is typically 5 times lower than $(0,4,4,6)$ and around an order of magnitude lower than $(0,2,8)$.}  We also monitor $Q_4$, which is sensitive to structures with local cubic symmetry.  We also calculated the locally averaged, bond-order parameters $\bar{Q}_l$ defined by Lechner and Dellago~\cite{Lechner_Dellago_2008}.   We found that these quantities show qualitatively similar trends as the standard BOO parameters defined above. However, the analysis of $\bar{Q}_l$ is complicated by the different symmetry of the local structures around A and B particles in the KA mixture.

To complement our study of crystal-like local order, we perform a common neighbor analysis of the system~\cite{Honeycutt_Andersen_1987}.  Within this approach, bonds between pairs of particles are assigned a triplet of integers $\{k,l,m\}$  that characterizes the connectivity between neighboring particles.  In particular, we keep track of the fraction $f_{\{1,4,2\}}$ of bonded pairs of particles that have four mutual neighbors ($l=4$) and are such that mutual neighbors share exactly two bonds between each other $(m=2)$.  High concentrations of $\{1,4,2\}$ bonds are indicative of fcc/hcp crystalline order. As in previous work~\cite{Hedges2009}, our criterion to discard trajectories generated by TPS with too high crystalline order is based on such $\{1,4,2\}$ bonds, whose average concentration must not exceed 8\% over the whole trajectory. We note that the concentration of such bonds is higher in inherent structures than in ``instantaneous'' configurations.

\begin{figure}[t]
  \centering
  \includegraphics[width=0.75\linewidth]{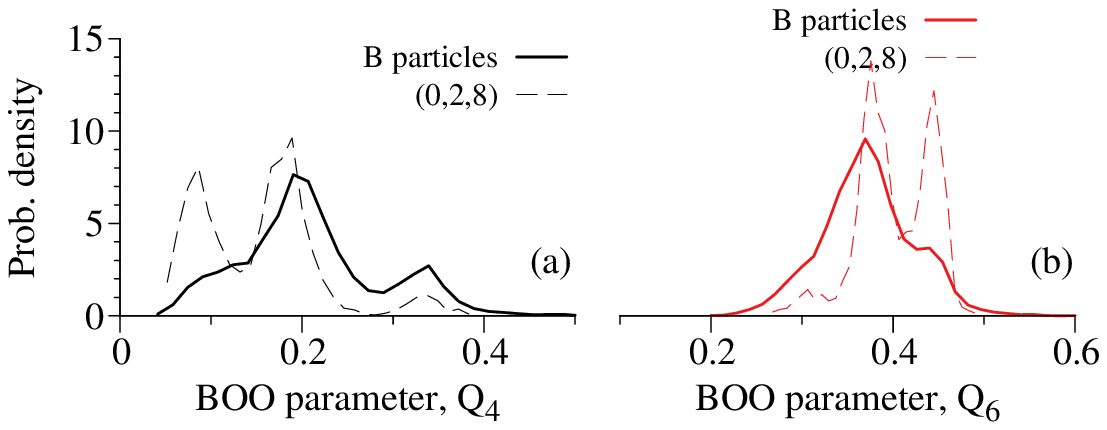} 
  \includegraphics[width=0.75\linewidth]{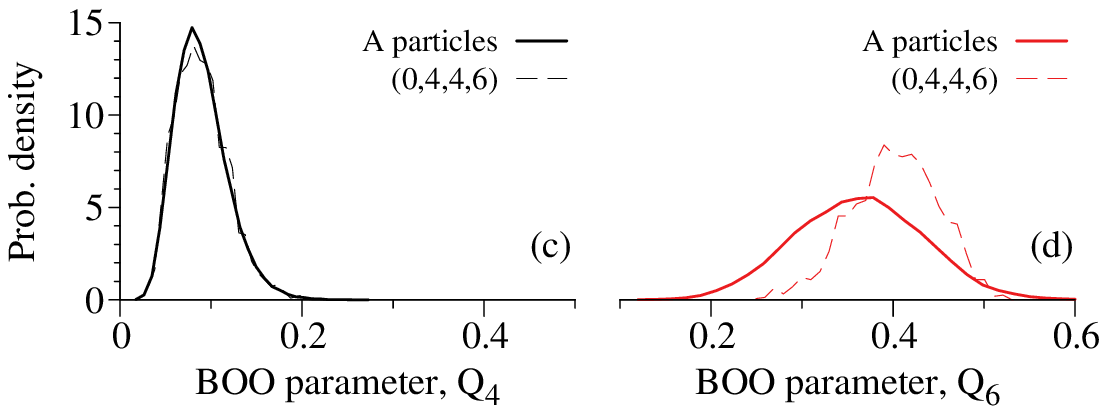}
  \caption{\label{fig:boo}Distribution of BOO parameters $Q_4$ [(a) and (c)] and $Q_6$ [(b) and (d)] from equilibrium configurations at $T=0.45$. Contributions due to $B$- and $A$-particles are reported separately in panels (a),(b) and (c),(d), respectively. Dashed lines are partial distributions filtered according to the indicated Voronoi signature around the particle.}
\end{figure}

Finally, we comment on the sensitivity of our local structure measurements to the neighbors' definition.
In Figure~\ref{fig:boo} we show the equilibrium distribution of $Q_4$ and $Q_6$ at $T=0.45$, split into contributions due to A and B particles. The distribution of $Q_4$ around $B$ particles is broad and characterized by distinct peaks. We think these peaks are an artifact of using a fixed nearest neighbor distance, which does not always account for the relevant local connectivity  (see Figure~\ref{fig:snaps}a and \ref{fig:snaps}b).
We found that these peaks are absent when using the Voronoi-based definition of nearest neighbors. Figure~\ref{fig:boo} shows that (0,2,8) cells are associated to larger (smaller) values of $Q_6$ and ($Q_4$). Analysis of distributions around A-particles shows that (0,4,4,6) structures are associated to systematically larger values of $Q_6$, as expected. These trends are robust and hold irrespective of neighbors' definition.
Finally, we found that the neighbors' definition significantly affects the statistics of CNA bonds. In particular, the number of (1,4,2) bonds varies appreciably when switching to a Voronoi-based definition. These discrepancies do not affect qualitatively our main conclusions, but they highlight the difficulty of an unambiguous identification of local structural motifs in glassy systems.

\section{Results}
\label{sec:results} 
The results presented in this work are of two types.  We first consider averaged structural measures in the inactive state, and we compare with analogous results for equilibrium states at various temperatures.  These results show that, on average, inactive states have energies and LFS populations consistent with equilibrium states at $T\approx 0.435$.  However, the inactive states differ from the equilibrium ones in that their local packing has somewhat more bond-orientational order, and the CNA analysis indicates a slightly higher proportion of $\{1,4,2\}$ bonds, which are associated with crystalline packing. This indicates that the local packing of particles in the inactive state differs from packing in low-temperature equilibrium states, although we emphasise that these structural differences are very subtle and 
the system does not display extended regions of crystalline order.

These results illustrate which kinds of local structure appear in typical inactive states, but they cannot detect whether there is any causal connection between these structures and the stability of these states, nor whether a measurement of structure allows the dynamical properties of a configuration to be predicted.  The second part of our analysis addresses this question by considering the fluctuations within the inactive state.   The local structural measures that we compute have significant fluctuations in this state, and are only weakly correlated with the energies and the kinetic stabilities of the different inactive configurations.  For example, while the average number of LFS is larger in the inactive state, there are still many inactive configurations with low energies but whose LFS populations are small, comparable to high-temperature equilibrium states. Consistent with previous studies~\cite{Hocky2014,Jack2014}, this indicates that the correlations between LFS concentrations and dynamics in the KA model are not strong enough to make predictions of dynamical properties. 

\subsection{{Averaged structural measurements in equilibrium and inactive states}}
\begin{figure}[t]
  \centering
\includegraphics[width=0.7\linewidth]{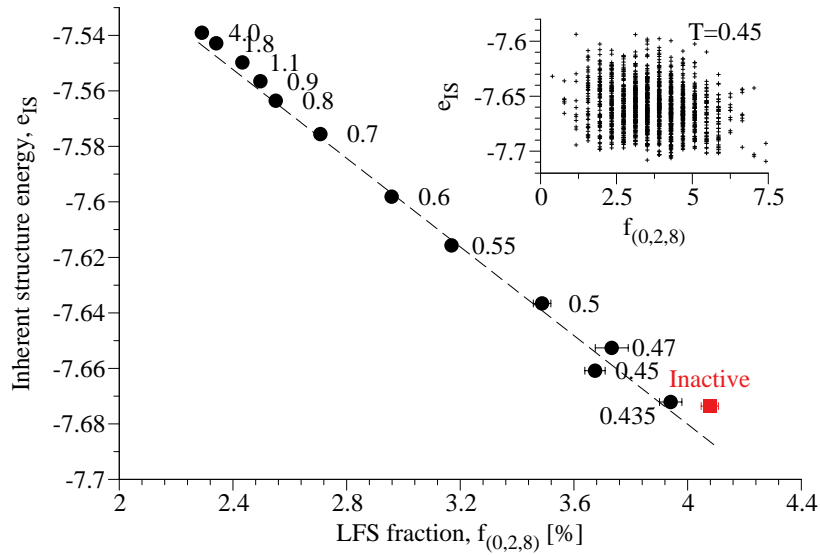}
\caption{\label{fig:eis_lps}
  Average inherent structure energy $e_{IS}$ plotted against the average LFS concentration $\lfs$, at equilibrium (circles) and in the inactive state at $T=0.6, s=s^*$ (red square).  The temperatures of the equilibrium samples are given by the labels.  Error bars indicate the numerical uncertainties on these average values (standard error).  
Inset: scatter plot of $e_{IS}$ against ${\lfs}$ for individual inherent structures sampled at equilibrium at $T=0.45$. The correlation between $e_{IS}$ and ${\lfs}$ is weak ($R=-0.16$).
}
\end{figure}

Figure~\ref{fig:eis_lps} shows the evolution of the preferred local order as the temperature of the KA mixture is reduced.  The equilibrium values $\ave{\eis}$ and $\ave{\lfs}$ are plotted one against the other by using temperature as an implicit parameter.  We emphasize that $f_{028}$ is evaluated using the inherent structures of the system, as are the other structural quantities (bond order, CNA) considered in the following.  The relationship between the average energy and the fraction of LFS is approximately linear.
Thus, the increased stability of inherent structures below the onset of slow dynamics, $T_O\approx 1$, is correlated with the growing fraction of LFS~\cite{Coslovich-PRE2011}.  Superficially, these data also suggest that increasing the number of $(0,2,8)$ cells present in the system bears a fixed energy cost, which is given by the slope of the straight line in Figure~\ref{fig:eis_lps}.  This would imply that individual structures are essentially independent in this temperature regime, in agreement with the small associated correlation lengths~\cite{Malins_Eggers_Tanaka_Royall_2014}.  Note, however, that the fluctuations of $\lfs$ and inherent structure energies do not correlate strongly on a sample-to-sample basis, as shown in inset of Figure~\ref{fig:eis_lps} for $T=0.45$: we return to this point in Section~\ref{subsec:cor}. Throughout this work $R$ indicates Pearson's correlation coefficient, which is defined for observables $A$ and $B$ as $R={\langle \delta A \delta B\rangle/\sqrt{\langle \delta A^2\rangle\langle \delta B^2\rangle}}$, with $\delta A = A-\langle A \rangle$, and similarly for $\delta B$.

\begin{figure}[t]
  \centering
\includegraphics[width=0.85\linewidth]{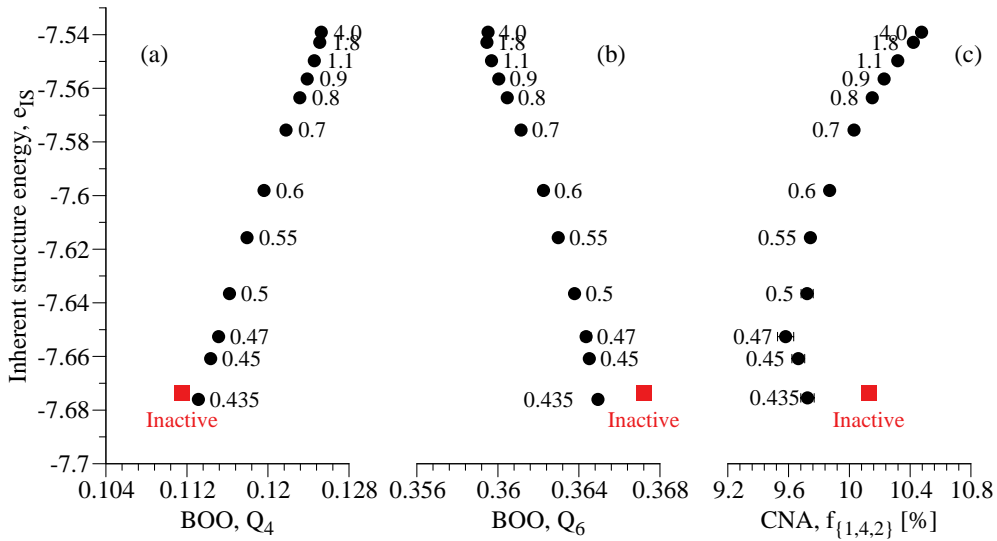}
\caption{\label{fig:eis_boo}
Average inherent structure energy $e_{IS}$ plotted against average BOO parameters (a) $Q_4$ (b) $Q_6$ and (c) concentration of $\{1,4,2\}$ bonds for inherent structures sampled at equilibrium (circles) and in the inactive state at $T=0.6$, $s=s^*$ (red square). The temperatures of the equilibrium samples are given by the labels.}
\end{figure}

Also shown in Figure~\ref{fig:eis_lps} is the average behavior of the inactive state.  As noted previously~\cite{Hedges2009,Jack2011}, this state is lower in energy than the active state with which it coexists, whose structural properties are close to equilibrium at temperature $T=0.6$.  The inactive state also includes more LFS~\cite{Speck2012a}, and appears to lie on the extrapolation of the equilibrium line, so its structure appears similar to that of equilibrium states at $T\approx 0.435$. 
However, there are also subtle differences in local structure between the inactive state and low-temperature equilibrium states.  This is illustrated in Figure~\ref{fig:eis_boo} where we show results for the average BOO parameters $Q_4$ and $Q_6$, and for $\cna$ bonds in the CNA analysis. $Q_6$ and $f_{\{1,4,2\}}$ are measures of structural order that have large values in fcc and hcp crystals. 
We see that inactive states have slightly more pronounced 6-fold bond-orientational order and as well as higher concentration of $\{1,4,2\}$ bonds than equilibrium configurations at a similar depth in the energy landscape.
We emphasize that these measurements only quantify \emph{local} packing and do not imply that the system is exhibiting long-range, crystalline order. Also, note that while the biasing field $s$ has been found to induce crystallization in this system, leading to long-ranged order~\cite{Hedges2009}, our transition path sampling method rejects trajectories in which crystalline order grows too large  (see Section~\ref{sec:bias}), ensuring that our results include only amorphous inactive states.  

Comparing with previous work~\cite{Jack2011}, we note that while inactive states were compared with low-temperature equilibrium states in that work, their smaller system size ($N=150$) meant that reliable results for equilibrium states could not be obtained below $T\approx 0.47$, due to crystallization.  Here, the use of larger systems ($N=256$) allows equilibration at temperatures as low as $T=0.435$.  However, the properties of the inactive states do depend on the system size: based on their IS energies, the inactive states ($N=150$) considered in~\cite{Jack2011} seem to correspond with equilibrium states close to $T\approx 0.4$, significantly lower than we find here, for $N=256$.  The larger systems considered here are also less kinetically stable (fitting the relaxation of the average energy during melting of the inactive state, we obtain an average $t_{\rm melt} \approx 140\Delta t$, in contrast to a value of $290\Delta t$ from~[37]).  The origins of these finite-size effects are not clear to us -- it would be interesting to investigate these further.  It is also useful to compare our results with~[46], where two different biasing parameters were used, leading to inactive states with different structures.  Our results are consistent with theirs for the case where their bias is purely dynamical; they also used a bias which couples to the local structure of the system, in which case the structure of the inactive state is (unsurprisingly) rather different.

\subsection{Correlation between structural and dynamical properties within the inactive state}
\label{subsec:cor}

\begin{figure}[t]
  \centering
  \begin{tabular}{cc}
    \includegraphics[width=0.35\linewidth]{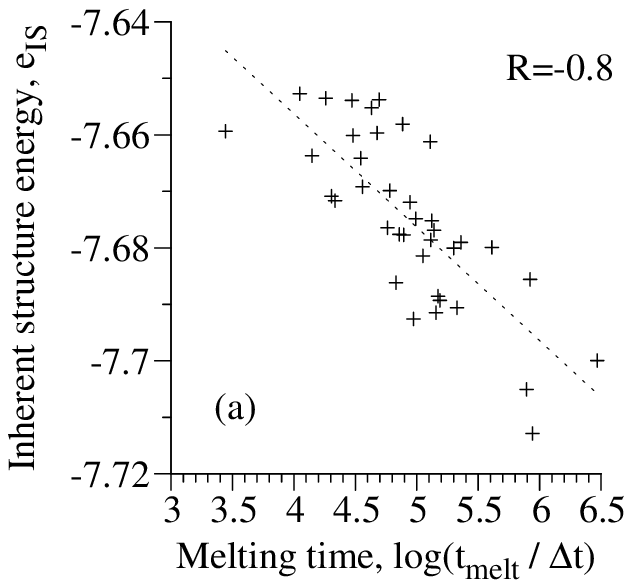} 
    \includegraphics[width=0.35\linewidth]{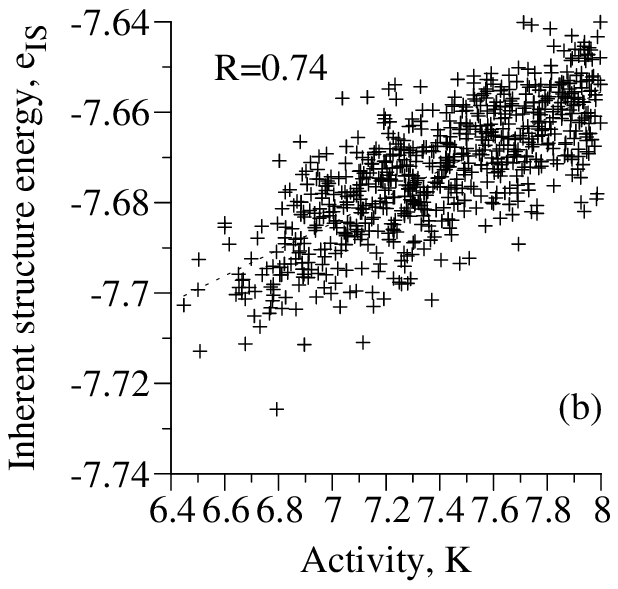} &
  \end{tabular}
\caption{\label{fig:corr1} Average inherent structure energies of inactive trajectories correlate well with (a) melting times $t_\mathrm{melt}$ and (b) activity $K$. In panel (a), $\log$ denotes the natural logarithm. The corresponding Pearson correlation coefficients $R$ are indicated in the figure. Note that the scatter plot in (a) is based on a representative subset of inactive trajectories. Averages of $\eis$ are evaluated over the central part of the trajectory, as described in Section~\ref{sec:bias}.}
\end{figure}

The inactive state considered here is at (dynamical) phase coexistence with a near-equilibrium active state at $T=0.6$.  The active and inactive states differ strongly in energy and in their structure, and these differences show that structural and dynamical properties of these states are correlated. However, establishing a causal relationship between these properties is much more challenging. We address this issue by analyzing fluctuations \textit{within the inactive state}, which allows us to demonstrate that some of the correlations found so far are not strong enough to form the basis of a causal link. To suppress some fast ``intra-state'' fluctuations, we average the structural properties of interest over the central part of the inactive trajectories, as described in Section~\ref{sec:bias}. 

Perhaps the most striking dynamical property of configurations from the inactive state is their kinetic stability:  if inactive states are allowed to evolve under their natural (unbiased, $s=0$) dynamics, they take a long time to relax (melt) back to equilibrium~\cite{Jack2011,Speck2012b}.  We measured the time $t_{\rm melt}$ associated with this relaxation process, as described in Section~\ref{sec:melt}.  Figure~\ref{fig:corr1} shows a scatter plot of  $\log(t_{\rm melt})$, against the structural quantity $e_{IS}$ for a representative set of inactive trajectories.  One sees a significant correlation: among all inactive trajectories, those with lower inherent structure energies are dynamically more stable.  Figure~\ref{fig:corr1}b shows a similarly strong correlation between the inherent structure energy and the dynamical activity of the individual trajectories. 

\begin{figure}[t]
  \centering
  \begin{tabular}{cc}
    \includegraphics[width=0.35\linewidth]{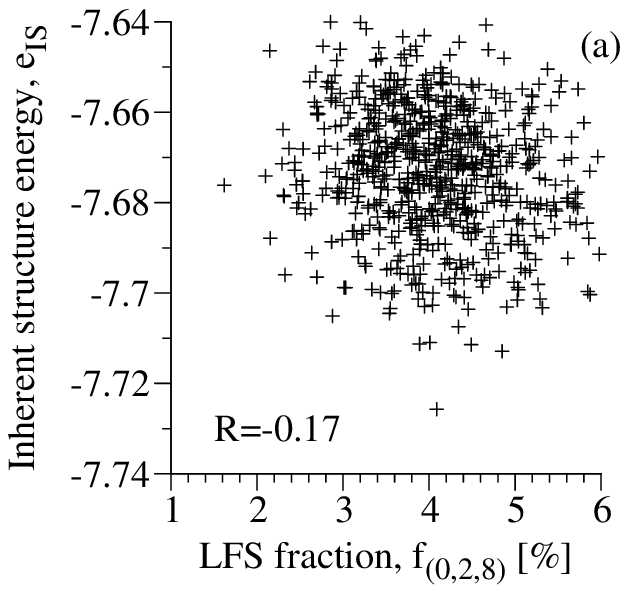} &
    \includegraphics[width=0.35\linewidth]{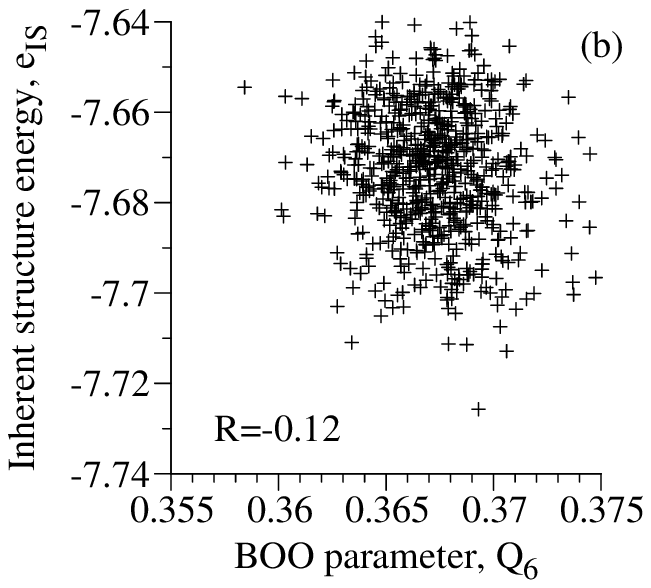} 
  \end{tabular}
\caption{\label{fig:corr2} Measures of local order, such as (a) the average fraction of LFS and (b) the average bond-orientational order parameter $Q_6$, do not correlate well with the average inherent structure energy $e_{IS}$ of inactive trajectories. The corresponding correlation coefficients $R$ are indicated in the figure. Averages are evaluated over the central part of the trajectory, as described in Section~\ref{sec:bias}.}
\end{figure}

By contrast, Figure~\ref{fig:corr2} shows similar scatter plots between $e_{IS}$ and local structural measures.  Here the correlations are much weaker: states with large numbers of LFS have lower energy on average (recall Figure~\ref{fig:eis_lps}), but measuring the number of LFS in an inactive trajectory provides very little information about its IS energy.  We emphasize that these correlations are as weak as those measured at the level of single inactive configurations, as shown in the inset of Figure~\ref{fig:eis_lps}. 

The correlations between several measured properties of inactive states are summarized in Table~\ref{tab:cor}.  We found that both Pearson and Spearman (i.e. rank-ordered) correlation coefficients provide similar results, with the former being systematically higher by a few percent.   
Overall, individual measures of local structure, be they associated to LFS or to local crystalline order, correlate weakly to inherent structure energy.
Somewhat surprisingly, the average bond-orientational parameter $Q_4$ has the strongest correlation among all local observables we investigated~\footnote{Evaluating the $Q_4$ parameter using the Voronoi neighbors (instead of neighbours based on a bond-length cutoff) and using an additional limitation of at most 12 neighbors, the correlation coefficient with $t_{melt}$ is reduced to 0.4 and the one with $\eis$ to 0.19.}, which suggests that $Q_4$ may capture some relevant, and so far unnoticed, features of the structure of this model. It may be interesting to investigate this point further in future work.

We emphasise that the analyses of local structure leading to Table~\ref{tab:cor} are all based on inherent structures, in order to reduce the thermal distortion of the local structure.  We also evaluated these correlations based on instantaneous configurations, in which case LFS concentrations correlate somewhat more strongly with instantaneous energies ($R \approx 0.29$). Correlations with melting times remain, however, practically absent ($|R|\approx 0.1$).

We also analyzed some additional local order parameters that have been used recently to characterize the structure of the KA mixture.  We investigated the role of the chemical composition of LFS structures~\cite{turci}, focusing on (0,2,8) Voronoi cells having two B-particles along the symmetry axis of the prism. This structure, depicted in Figure~\ref{fig:snaps}b, can expected to be a favorable motif at low temperature~\cite{turci}.
We found the correlation between the fraction of such structures and $\eis$ to be even lower than for the full statistics of (0,2,8) cells.  Similar weak correlations are found with the fraction of $A$ particles fully surrounded by $A$ particles, which is a simple measure of crystalline local order in the KA mixture~\cite{Hedges2009}.
Finally, we note that measurements of $f_{(0,2,8)}$ obtained
using the modified Voronoi tessellation described in Section~\ref{sec:structure} are 95\% correlated to those reported
here and provide similar, weak correlation coefficients with energy and activity (both equal to 0.21).  

We emphasize that while the correlations in Figure~\ref{fig:corr2} are very weak, they are non-zero. 
We note that similar weak correlations also hold in equilibrium states, as may be deduced from Figure~\ref{fig:eis_lps}.  This means that if one averages $\eis$ and $\lfs$ over several configurations, the fluctuations in these quantities decrease, and their correlation becomes apparent.  This effect can be seen, for example, by averaging over time windows that are long enough that a single trajectory visits many different structures.  In that case, a correlation appears between time-averaged values of $\eis$ and $\lfs$~\cite{Speck2012b}, but this does not mean that measurement of a configuration's local structure provides predictive information for the dynamical properties of the system.

\begin{table}[t]
  \caption{\label{tab:cor} Pearson correlation coefficients between observables measured in the inactive state.
    Correlations are measured using averages over the central part of the TPS trajectory, , as described in Section~\ref{sec:bias}, except for $K$, which is defined over the full trajectory, and $(t_{\rm melt},\epsilon^*,\chi^*) $, which depend only on the configuration at the mid-point of the trajectory ($t=\tobs/2$).    
    Correlation coefficients for the melting time are evaluated using $\log{t_{\rm m}}$.
  Measurements performed using a representative subset of inactive trajectories carry only one significant digit and are typeset in italic.
  $R$ values larger than 0.5 are indicated by bold text.}
\lineup
\vskip1ex
 \begin{indented}
   \item[] 
    \hskip-5ex
   \begin{tabular}{@{}lrrrrrrrrrrrrr}
      \br
      		&$e_{IS}$	&$K$	&$f_{(0,2,8)}$	&$f_{(0,4,4,6)}$	&$Q_6$	&$Q_4$	&$f_{\{1,4,2\}}$	&$t_{melt}$	&$\epsilon^*$	&$\chi^*$\\
\hline
$e_{IS}$	 &$-$	&\textbf{0.75}	& \textrm{-0.18}	& \textrm{-0.06}	& \textrm{-0.12}	& \textrm{0.33}	& \textrm{-0.01}	& \textbf{-0.8}	& \textbf{0.5}	& \textbf{-0.6}	\\
$K$	 &\textbf{0.75}	& $-$	&\textrm{-0.21}	& \textrm{-0.05}	& \textrm{-0.20}	& \textrm{0.36}	& \textrm{-0.09}	& \textbf{-0.7}	& \textit{0.4}	& \textit{-0.4}	\\
$f_{(0,2,8)}$	 &\textrm{-0.18}	& \textrm{-0.21}	& $-$	&\textrm{-0.22}	& \textrm{0.17}	& \textrm{-0.32}	& \textrm{-0.06}	& \textit{-0.1}	& \textit{-0.1}	& \textit{-0.1}	\\
$f_{(0,4,4,6)}$	 &\textrm{-0.06}	& \textrm{-0.05}	& \textrm{-0.22}	& $-$	&\textrm{0.21}	& \textrm{-0.09}	& \textrm{0.23}	& \textit{0.0}	& \textit{-0.2}	& \textit{-0.0}	\\
$Q_6$	 &\textrm{-0.12}	& \textrm{-0.20}	& \textrm{0.17}	& \textrm{0.21}	& $-$	&\textrm{-0.25}	& \textrm{0.26}	& \textit{0.2}	& \textit{-0.2}	& \textit{0.0}	\\
$Q_4$	 &\textrm{0.33}	& \textrm{0.36}	& \textrm{-0.32}	& \textrm{-0.09}	& \textrm{-0.25}	& $-$	&\textrm{0.30}	& \textbf{-0.5}	& \textit{0.3}	& \textit{-0.2}	\\
$f_{\{1,4,2\}}$	 &\textrm{-0.01}	& \textrm{-0.09}	& \textrm{-0.06}	& \textrm{0.23}	& \textrm{0.26}	& \textrm{0.30}	& $-$	&\textit{0.2}	& \textit{-0.2}	& \textit{0.1}	\\
$t_{melt}$	 &\textbf{-0.8}	& \textbf{-0.7}	& \textit{-0.1}	& \textit{0.0}	& \textit{0.2}	& \textbf{-0.5}	& \textit{0.2}	& $-$	&\textbf{-0.7}	& \textbf{0.7}	\\
$\epsilon^*$	 &\textbf{0.5}	& \textit{0.4}	& \textit{-0.1}	& \textit{-0.2}	& \textit{-0.2}	& \textit{0.3}	& \textit{-0.2}	& \textbf{-0.7}	& $-$	&\textbf{-0.8}	\\
$\chi^*$	 &\textbf{-0.6}	& \textit{-0.4}	& \textit{-0.1}	& \textit{-0.0}	& \textit{0.0}	& \textit{-0.2}	& \textit{0.1}	& \textbf{0.7}	& \textbf{-0.8}	& $-$\\

      \br
    \end{tabular}
  \end{indented}
\end{table}

From Table~\ref{tab:cor}, our conclusion is that standard measures of local order are not sufficient to capture the most important structural fluctuations observed in the inactive state. None of the individual structural measures give predictive information as to the IS energy or the kinetic stability. We see several
possible theoretical scenarios that are consistent with these data: (i) There are multiple local order measures whose fluctuations contribute to those of $\eis$, so that only the fluctuations of a more complex structural order parameter (for example, a sum of LFS and local crystalline order) can correlate to $\eis$. 
We also cannot rule out the possibility that different strategies to detect local structures, such as the topological cluster classification~\cite{malins_identification_2013}, lead to stronger correlations.
(ii) The dynamic quantities $K$ and $\log(t_m)$ are related to structural \textit{defects}, which increase $\eis$, but which are not captured by any of the local order measures. The latter are primarily sensitive to low-energy stable regions. These structural defects are yet to be identified -- it would be interesting to investigate correlations between kinetic stability and the low density regions identified in~\cite{Jack2014}.  (iii) Some more complex forms of order are responsible for the stability of the inactive state, for example soft spots~\cite{WidmerCooper-NatPhys2008,Jack2011}, medium ranged-order~\cite{Tanaka_2012,Ma_2015}, or point-to-set correlations~\cite{Berthier-static-PRE2012,Fullerton2014,BerthierJack2015}.

To investigate further this last scenario of strong point-to-set correlations and amorphous order, we revisit the results of~\cite{BerthierJack2015}.  In that work, a field $\epsilon$ was introduced, which biases one copy of a system to be similar to a second copy, which is fixed in an inactive configuration $\CC_0$.  The theoretical prediction~\cite{Franz1997} is that if the temperature is sufficiently low and the inactive state is very stable, one should observe a first-order phase transition as a function of the bias $\epsilon$.  The order parameter for this phase transition is the overlap $Q$, which measures the similarity between the two copies of the system, with $Q=1$ if the configurations are identical and $Q\approx0$ if they are uncorrelated.  

We consider configurations $\CC_0$ taken from the inactive state.  Characterizing the expected phase transition requires averaging over many different $\CC_0$, but we consider each $\CC_0$ separately here, in order to consider fluctuations within the inactive state.  For each fixed $\CC_0$, a sharp crossover in $Q$ was observed~\cite{BerthierJack2015}, as the bias $\epsilon$ is increased.  The value of the bias at which this crossover takes place is $\epsilon^* = \epsilon^*(\CC_0)$.  At the crossover, one may also measure a susceptibility $\chi^*$, which is related to the variance of the order parameter as $\chi^* = N\langle (\delta Q)^2 \rangle$, where the average is taken at fixed $\CC_0$, with $\epsilon=\epsilon^*$.  One also has $\chi^* = (d/d\epsilon) \langle Q \rangle$.

\begin{figure}[t]
\centering
  \begin{tabular}{cc}
    \includegraphics[width=0.35\linewidth]{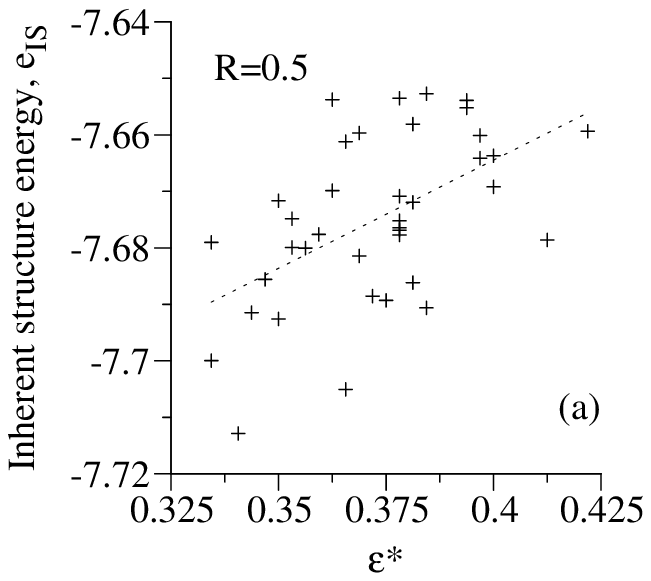} &
    \includegraphics[width=0.35\linewidth]{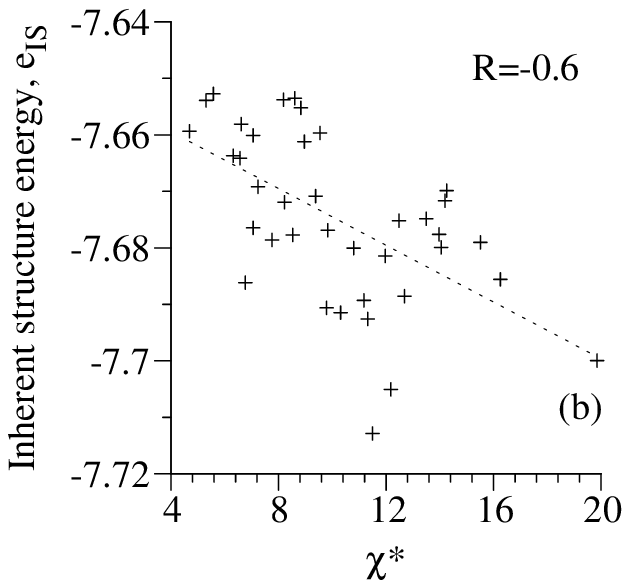} 
  \end{tabular}
\caption{\label{fig:corr_eps} Correlation between average inherent structure energy and (a) $\epsilon^*$ and (b) the maximum of overlap susceptibility $\chi^*$. Note that only the central configurations from the inactive trajectories are used in the calculation of $\epsilon^*$ and $\chi^*$.}
\end{figure}

The physical meaning of $\epsilon^*$ and $\chi^*$ is as follows.  If $\epsilon^*$ is small, this means that only a small bias is required in order to localize the system in the same metastable state as configuration $\CC_0$.  This localization has a free-energy cost associated with the loss of entropy of the parent fluid (here the active state)~\cite{BerthierCoslo2014}; there is also a free energy gain since the energy of $\CC_0$ is lower than that of the parent fluid.  The net free-energy cost for localization must be compensated by the bias $\epsilon$, so one expects $\epsilon^*$ to be small if $\CC_0$ is a low-energy stable configuration.  This expectation is confirmed by Figure~\ref{fig:corr_eps}a, which shows a good correlation between $\epsilon^*(\CC_0)$ and the IS energy of $\CC_0$, with lower energy corresponding to smaller $\epsilon^*$.  The meaning of the susceptibility $\chi^*$ is more subtle: it measures how sharp is the transition between the parent fluid and the system localized in the inactive metastable state.  One expects the transition to be sharp (and hence $\chi^*$ large) in cases where there is a significant interfacial cost between active and inactive states, leading to bistable behavior for the overlap~\cite{JackGarrahan2016}.  These large interfacial costs are also expected to be correlated with large time scales for relaxation, following the arguments of~\cite{Bouchaud2004}. Figure~\ref{fig:corr_eps}b shows that this expectation does indeed hold, in that large $\chi^*$ correlates to a good extent with low IS energy, which is itself strongly correlated with kinetic stability (Figure~\ref{fig:corr1}).

\section{Conclusions}

We have investigated the structure of inactive states obtained by a large deviation analysis, following~\cite{Hedges2009}.  The inactive states have low energy and high kinetic stability.  These states are prepared at dynamical phase coexistence with an equilibrium state at $T=0.6$ but, on average, their local structures are similar to those of equilibrium states at $T\approx 0.435$.  However, CNA and BOO measurements reveal subtle differences between inactive and low-temperature equilibrium states.  We also find that the inactive state has quite large fluctuations in its local structure.  For example, the concentration of LFS ranges from $2-6\%$ (see Figure~\ref{fig:corr2}), even after averaging over fast fluctuations within the inactive state.  Correlations between the LFS concentration and the dynamical properties of the inactive state are weak, so LFS measurements alone cannot predict dynamical behavior in this system~\cite{Jack2014,Hocky2014}. Correlations between dynamics and other local order metrics are similarly weak. On the other hand, collective properties such as the IS energy and the coupled-replica measurements of~\cite{BerthierJack2015} do correlate with dynamical features, indicating that the coupling between structure and dynamics is strong, even if it may not be local.

The data underlying this publication are available at http://doi.org/10.15125/BATH-00209.

\ack
We thank Chris Fullerton for his help with the early stages of this work, and Ludovic Berthier for facilitating our analyses of $\epsilon^*$ and $\chi^*$.  We thank Paddy Royall and Thomas Speck for many useful discussions on the subject of local structure. RLJ thanks the Engineering and Physical Sciences Research Council (EPSRC) for financial support through grant EP/I003797/1.

\section{Bibliography}

\providecommand{\newblock}{}

\end{document}